\documentstyle[floats,twocolumn,amssymb,aps,epsfig,pre]{revtex}
\draft

\begin{document}
\wideabs{ \title{Force distributions in  3D granular assemblies:\\
Effects of packing order and inter-particle friction} \author{Daniel
L. Blair, Nathan W. Mueggenburg, Adam H. Marshall, Heinrich M. Jaeger,
Sidney R. Nagel} \address{The James Franck Institute and Department of
Physics\\ The University of Chicago \\ 5640 S. Ellis Ave. Chicago, IL
60637} \date{Received 20 September 2000} \maketitle

\begin{abstract}
We present a systematic investigation of the distribution of normal
forces at the boundaries of static packings of spheres.  A new method
for the efficient construction of large hexagonal-close-packed
crystals is introduced and used to study the effect of spatial
ordering on the distribution of forces.  Under uniaxial compression we
find that the form for the probability distribution of normal forces
between particles does not depend strongly on crystallinity or
inter-particle friction.  In all cases the distribution decays
exponentially at large forces and shows a plateau or possibly a small peak near
the average force but does not tend to zero at small forces.
\end{abstract}

\pacs{PACS numbers:  81.05.Rm, 45.70.-n, 83.70.Fn, 45.70.Cc}

}

\section{Introduction}
In ordinary solids and confined fluids, uniformly applied loads are
distributed homogeneously throughout the material.  This is not the case
for granular materials, which are collections of discrete non-cohesive
macroscopic particles~\cite{jaeger96}.  For such systems, stresses are
distributed in a highly inhomogeneous manner, along networks containing
the largest inter-particle
forces~\cite{dantu57,dantu67,howell97,ammi87,liu95}.  These ``force
chains'' support most of the external load, effectively shielding large
regions of the material.  The probability distribution, $P(F)$, of normal
forces, $F$, between neighboring particles provides a quantitative way
of analyzing these stresses~\cite{liu95}.  $P(F)$ in granular media
exhibits characteristic differences from what would be expected for
ordinary solids or fluids.

Recent experiments have measured $P(F)$ at the container boundaries of
granular media~\cite{liu95,mueth98,lovoll99,makse00,baxter97}.  When the
applied loads are small enough not to cause significant particle
deformations two key features of $P(F)$ have emerged: first, $P(F)$
exhibits an abundance of small force values, even far below the mean
force, $\overline F$; and second, for forces above $\overline F$, $P(F)$
decays with a characteristic, exponential dependence.  This unusual shape
of $P(F)$ has wide-ranging implications.  Compared to a Gaussian profile,
the exponential tail indicates a significantly higher probability of an
individual contact force greatly exceeding the mean force.  In
applications with a fixed yield strength, this corresponds to higher
chances for material failure.  Perhaps even more intriguing is the
non-zero value for $P(F)$ as $F$ approaches zero.  Investigations
concerning the structure of force chains have suggested that this
abundance of low forces could be caused by arching
effects~\cite{claudin97,luding97,duran98,peraltafabi99}.  The shape of
$P(F)$ at low forces has also been associated with glassy
behavior~\cite{ohern}.

As far as the precise functional form for $P(F)$ is concerned,
however, there is currently no clear consensus.  Several fitting forms
have been proposed, which differ most widely in their predictions for
$P(F)$ at low forces, ranging from $P(F)\to 0$~\cite{coppersmith96} to
$P(F)\to \infty$~\cite{claudin97,radjai96,radjai97} as $F$ approaches zero.
Because of the obvious discrepancies not only among some of these model
predictions, but also between certain model predictions and the
available experimental data, it is important to test the robustness of
the results with regard to variations in the bead parameters, the
packing construction, and the loading conditions.  Experimentally, the
issue of whether $P(F)$ depends on inter-particle friction in any
significant way and whether disordered packings behave differently
from highly ordered, crystalline configurations has not been tested.
Clearly, a perfect, infinite crystal composed of identical grains
would result in a delta function for $P(F)$ since all contact forces
are the same in this case.  However, it is unknown how the shape of
$P(F)$ is modified by the small amounts of disorder present in real
granular crystals.  Also it is not known how inter-particle friction
influences the distribution of forces.

Here we address these issues with a systematic experimental investigation
of the effects on $P(F)$ of packing order as well as inter-particle
friction.  Our results confirm the robust character of the probability
distribution of forces.  We find that $P(F)$ is essentially independent of
changes in the particle arrangement from amorphous to crystalline and
changes in the particle surface from smooth to rough.  In particular, we
observe an exponential decay over more than 2.5 decades for forces larger
than the mean.  Below the mean force, our data are consistent with either
a flat distribution or a small peak centered close to $\overline F$.

The paper is organized as follows.  After a review of currently available
theoretical model predictions and simulation results, we introduce our
experimental methods, including details of the carbon paper technique used
to measure normal forces and a discussion of the experimental
uncertainties.  We then present results for $P(F)$ obtained from
crystalline and amorphous granular packings using both smooth and rough
beads.  We also show how $P(F)$ evolves with system depth. A final
section discusses these results in light of recent experimental and
theoretical work on granular and other jammed systems.

\section{Background}

The original $q$-model, proposed by Coppersmith {\em et
al}.~\cite{liu95,coppersmith96} captures the dominant behavior of $P(F)$
at large $F$:  $P(F/\overline F)\propto \exp (-d\,F/\overline F)$, where
$d$ is a positive constant.  This model was extended by Nguyen and
Coppersmith~\cite{nguyen99} and by Socolar and
Sexton~\cite{socolar98,sexton99}, and Claudin {\em et al}.\ have related
it to a diffusion equation with an additional randomly varying convection
term~\cite{claudin98}.  Despite this success, these models predict
power-law behavior for $P(F)$ at forces below the mean force, implying
$P(F)\to0$ as $F\to0$, which disagrees with experimental
findings~\cite{mueth98,lovoll99}.  Using contact dynamics computer
simulations, Radjai {\em et~al}.~\cite{radjai96,radjai97,radjai98} have measured $P(F)$ and fit it with a
slowly diverging power law below $\overline F$ and with a decaying
exponential at large forces.  This form
is qualitatively more consistent with experimental results~\cite{mueth98,lovoll99}.

Recently, the effects of particle deformations on the structure of force
chains has been considered.  Simulations by Makse {\em
et~al}.~\cite{makse00} show
evidence for a crossover in the shape of $P(F)$ from pure exponential to
Gaussian as the applied load is increased.  However,
experiments by L\o voll {\em et~al}.~\cite{lovoll99} with no external
force do not show
pure exponential behavior; rather, they are more
consistent with the results of Mueth {\em et~al}.~\cite{mueth98} at higher
deformations.  This is in agreement with simulations
by Thornton~\cite{thornton97,thornton98} and Nguyen and
Coppersmith~\cite{nguyen}, which observe a slow trend toward Gaussian
behavior at very large deformations with no evidence of a pure exponential
in the low-deformation regime.

Tkachenko and Witten have argued that frictional forces lead to elastic
behavior~\cite{tkachenko99}.  Simulations performed by Eloy and
Clement~\cite{eloy97} and by Thornton~\cite{thornton00} predict a
dependence of $P(F)$ on inter-particle friction, but this has yet to be
seen in experiment.

$P(F)$ has also been studied in simulations of supercooled liquids and
glasses~\cite{ohern}.  For a liquid with a strongly repulsive core the
tail of $P(F)$ is a decaying exponential at all temperatures.  As the
temperature of the liquid is lowered to the glass transition, $P(F)$
develops a small peak centered at $\overline F$ and is very
reminiscent of the experimental data on granular systems found by
Mueth {\em et al}.~\cite{mueth98}.

\section{Experimental Methods}

We have performed experiments on five different granular systems.  In
order to compare our results with the earlier work of Mueth {\em
et~al}.~\cite{mueth98}, we first examined an amorphous pack of smooth spherical glass beads.  We then
constructed both hexagonal-close-packed (HCP), and face-centered-cubic
(FCC) crystalline lattices of the spheres and repeated the experiment on
these highly ordered systems.  In addition, we examined both amorphous
and HCP arrangements of spheres with roughened surfaces.

\begin{figure}[tb]
\centerline{\epsfxsize= 7.0cm\epsfbox{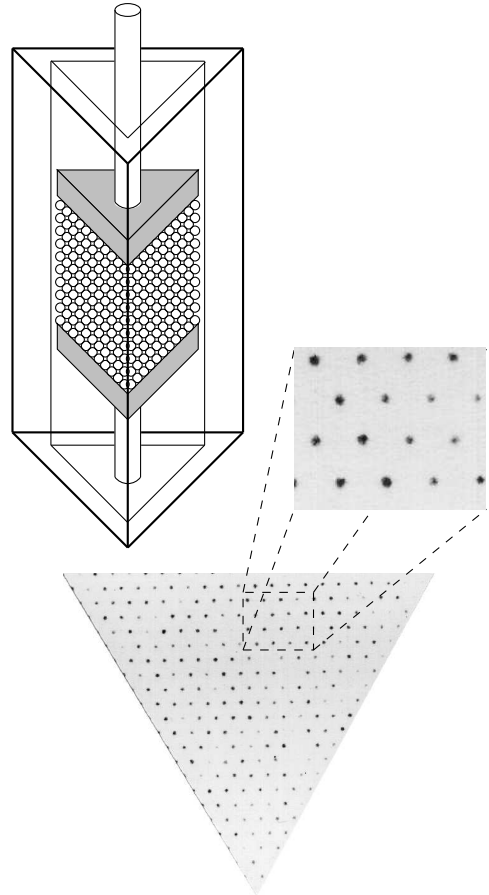}}
\vspace{.5cm}
\caption{ A schematic of the triangular cell used to contain both
crystalline and amorphous packs.  A force of 7600\,N was applied to
the rigidly constrained shaft on the top piston.  The cell is 53 bead
diameters on a side.  A portion of a scanned image of the imprints
left on the white paper by a crystal pack is shown with a close-up
highlighting the crystalline order.}
\label{fig1}
\end{figure}

The experiments were performed using approximately 70,000 soda lime
glass spheres with diameter $3.06\pm0.04\,\mathrm mm$.  We used an
acrylic container of equilateral triangular cross section with a side
length of 165\,mm, which is commensurate with our HCP crystals (See
Fig.~\ref{fig1}).  The top and bottom boundaries of the system were
thick, close-fitting acrylic pistons.  The height in layers, $h$, of
the HCP crystals ranged from 1 to 61, with the bulk of the data being
taken at $h=45$.  FCC crystals were studied with $h=10$.  The
amorphous arrangements had a filling height of 12\,cm, which
corresponds to approximately the same number of particles as in the
45-layer crystals.

\subsection{Crystal Packing}

It is difficult to construct three-dimensional macroscopic crystals of
granular materials.  Vibration methods for crystallization can
efficiently produce triangular close-packed layers~\cite{pouliquen97};
however these methods do not control the stacking order of the planes
which is necessary for three-dimensional crystal structures.  Also,
these methods typically result in large numbers of defects
(approximately 10\%) which can ruin long range order.  We have
developed a method to construct near-perfect crystals of glass spheres
in an efficient manner.  Our technique consists of four steps.  First,
one layer's worth of beads is placed upon a perforated metal sheet
which has small holes arranged in a triangular lattice with a spacing
equal to the bead diameter.  The array of holes, and therefore the
collection of beads, is commensurate with our cell.  Second, a vacuum
is established behind the perforated plate; this holds the beads in
place while the plate is manipulated.  Third, the arrangement is
inverted and lowered into the cell, which has been precisely machined
to accept a monolayer of crystallized beads.  The vacuum is then
removed to release the beads.  Fourth, a small number of residual
defects and grain boundaries are removed by hand before the process is
repeated to add successive layers.  By carefully monitoring the
layer-to-layer structure, we were able to create large, nearly perfect
HCP structures of arbitrary height with 1,431 and 1,378 particles in
alternating layers.  Upon construction, each 45-layer HCP crystal
contained 70,119 beads with fewer than 10 beads significantly out of
place, and typically these defects were confined to the edges of the
container.

HCP packings of rough spheres were created in the same manner.
Unfortunately, the etching process reduced the bead size by a few
micrometers, causing a noticable incommensurability between the bead
lattice and the cell.  This resulted in a larger number of defects and
less stability.

We also constructed a FCC crystal, but boundary effects made the
procedure considerably more difficult.  These effects were minimized
by rotating the orientation of the pack by thirty degrees.  A seed
layer was used at the bottom of the cell to arrange the pack in this
awkward position.  The resulting packing was incommensurate with our
cell and was highly unstable; for this reason only one small ($h=10$)
smooth FCC crystal was created.

\subsection{Rough Particles}

Force chains are present only when  grains are in contact.  This
suggests that the highly nonlinear frictional forces between particles
may play an important role in determining the network of force chains.
To understand the magnitude of these effects on $P(F)$, we have
varied $\mu$, the coefficient of static friction between grains.

\begin{figure}[tb]
\centerline{\epsfxsize=8.1cm\epsfbox{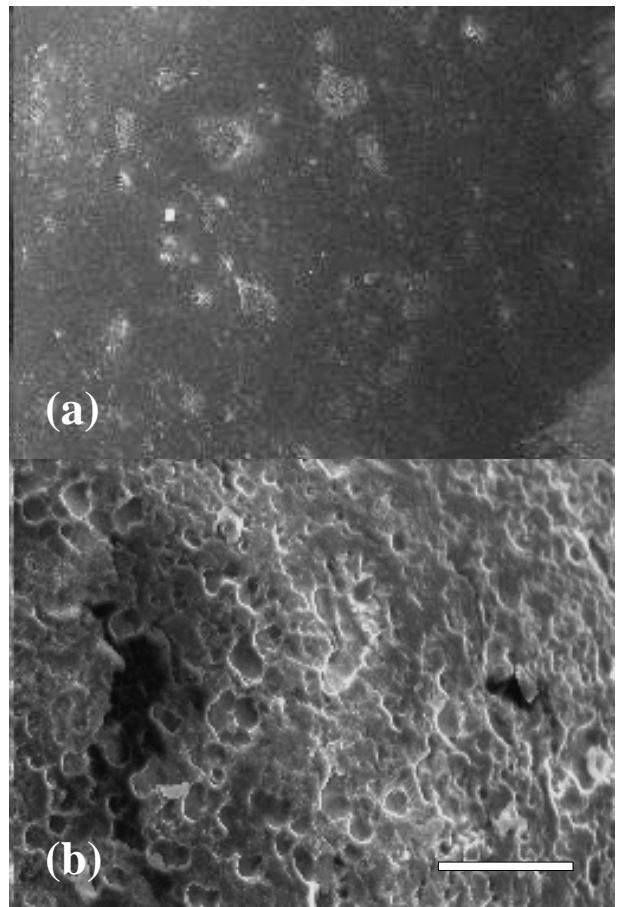}}
\caption{ Scanning electron microscope images of a) a smooth glass
bead and b) a glass bead after etching in hydrofluoric acid.  The
white bar is $100\,\mu{\mathrm m}$ in length.}
\label{fig2}
\end{figure}

\begin{figure}[tb]
\centerline{\epsfxsize=7.2cm\epsfbox{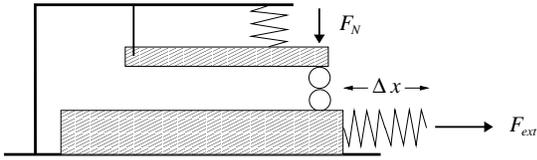}}
\vspace{.1in}
\caption{Schematic of the apparatus for measuring static friction
between two beads.  $F_N$ is the normal force applied to the beads,
and $\Delta x$ denotes the spring elongation required to move the
bottom bead which is attached to a low-friction sled.}
\label{fig3}
\end{figure}

A collection of glass spheres was etched in a hydrofluoric acid
solution (Ashland Chemical Inc.\ 7:1 Buffered Oxide Etch) for
approximately 2 hours.  The difference between the smooth beads and
the etched beads was quite apparent by visual inspection.
Figure~\ref{fig2} shows scanning electron microscope images of smooth
and etched beads.  We have quantified the increase in surface friction
between grains due to etching using the apparatus shown schematically
in Figure~\ref{fig3}.  The ratio of the coefficients of static
friction is characterized by the changes in spring length given by
\begin{equation}
\frac{\mu_{\mathit rough}}{\mu_{\mathit smooth}}=\frac{\Delta
x_{\mathit rough}-\Delta x_{\mathit sled}}{\Delta x_{\mathit
smooth}-\Delta x_{\mathit sled}}
\label{eq1}
\end{equation}
where the change in spring length, $\Delta x$,  is measured at the
point of bead movement, and $\Delta x_{\mathit sled}$ is included to
remove the contribution from friction of the sled with the base.  The
measured ratio was $3.0\pm0.3$\,.

\subsection{Analysis Techniques for Measuring Local Forces}

The experiments were performed as follows.  An external force of
7600\,N was applied to the top piston of the cell with a pneumatic
press, while the bottom piston rested  upon a fixed  floor.  Some
force was carried by the walls of the cell, but we determined that
this was typically less than 10\% of the applied force.  We measured
the normal forces at the upper and lower boundaries by placing carbon
paper and white paper between the granular material and the
pistons~\cite{materials}.  This allowed each grain at the boundaries
to press into the paper, leaving a mark whose size and intensity were
dependent upon the normal force on that
bead~\cite{liu95,mueth98,delyon90}.  A portion of a scanned image  of the
resulting imprints is shown in Figure~\ref{fig1}.  Roughly 2000 marks
from both the top and bottom boundaries were identified for each data
run.  We observed no significant deviations from circular marks, as
might be produced by tangential forces.

After each data run, each sheet of white paper was carefully removed and
digitized with a flat bed scanner.  These digital images were then
processed using image analysis software to find both the area and the
intensity of each mark.  The force on each bead was extracted from the
area and intensity of the corresponding feature using calibration curves
taken over the same force range.  To account for variations in the applied
load from run to run, the forces on the individual beads were divided by
the average force for each data run.  A histogram of the rescaled forces
was then averaged over typically 10-20 independent data runs and
normalized in order to represent the probability distribution of forces,
$P(f)$, where $f\equiv F/\overline F$.

\begin{figure}[tb]
\centerline{\epsfxsize=8.0cm\epsfbox{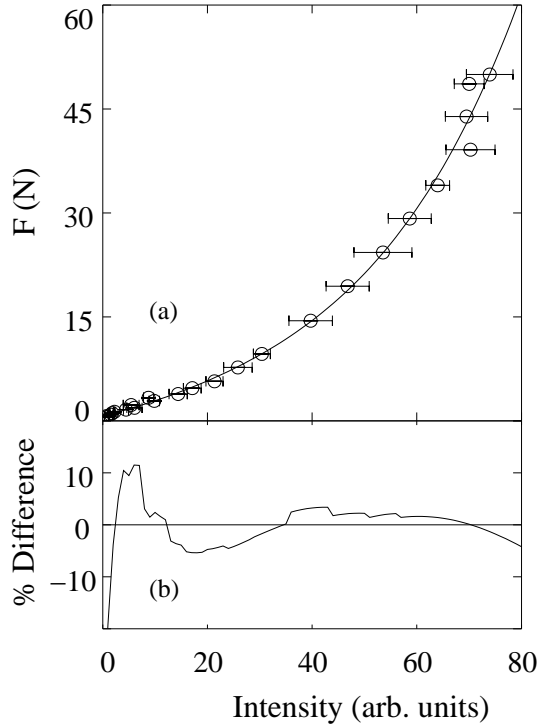}}
\caption{Calibration data.  a) A typical calibration set fit to a
global fourth-order polynomial (GFP).  b) Percent difference between
the fit in the top panel and a local sliding parabolic fit (LSP) to
the same data, given by $\mathrm 100\,(LSP-GFP)/GFP$.  While both
interpolations fit the data reasonably well, the differences below the
mean force ($\overline F\approx7\,\mathrm N$) shown in the bottom
panel can result in large changes to the calculated probability
distribution.}
\label{fig4}
\end{figure}

Calibration data were obtained by lowering a known mass  connected to
a single bead onto carbon paper covering white paper.  This system,
described in reference~\cite{mueth98}, consisted of a vertical slide with
impulse-absorbing springs; a pulley and cable system was added to
ensure that the bead was lowered slowly and monotonically in order to
further eliminate impulse forces.  Because of the potential
variability in carbon paper, each package of paper was calibrated
independently.  To interpolate a calibration curve from the discrete
set of calibration points we used two procedures: 1)~a global
fourth-order polynomial fit and 2)~a local sliding parabolic fit with
a window of eleven data points.  Figure~\ref{fig4}a shows a
representative set of calibration data together with the fitted
fourth-order polynomial.  The relationship between force and intensity
is monotonic but nonlinear.  We found that the force distributions
were very sensitive to the calibrations, especially in the low-force
regime.

\subsection{Systematic and Statistical Errors}

The uncertainties due to the calibrations are an inherent
characteristic of the carbon paper method and represent the largest
contribution to systematic errors.  The most significant problem is
the proper interpolation between calibration points.  As shown in
figure~\ref{fig4}b, the two interpolation methods can lead to
significant differences in the intensity-to-force conversion.  It is
not clear which of the two fitting methods gives the more accurate
calibrations, although we found the fourth order polynomial to produce
smoother data.  For the data shown below, we computed $P(f)$ using both
methods (see insets of figures \ref{fig5}~and~\ref{fig6}), and we
interpret the difference as an estimate of the systematic error
introduced by the calibration.

An additional complication arises from the presence of forces below our
threshold of detectability, which is approximately 0.7\,N.  It is
difficult to distinguish between undetectably small forces and missing
bead contacts, but the number of forces below threshold can be estimated
experimentally using the two-sided tape technique described by Mueth {\em
et al}.~\cite{mueth98}.  An approximate number of very small forces was
inserted into the force distributions to account for these undetectable
contacts.  Errors in the number of undetectable forces affect the
normalization and thus effectively rescale $f$.  Including this source
of error, we estimate the uncertainty in the exponential slope to be
approximately 5\%.

\section{Results}

We assembled a large number of amorphous and crystalline packs of both
smooth and rough glass beads and studied  them with the carbon paper
technique.  As detailed below, the force distributions exhibit
exponential behavior for $f>1$.  This result has been found previously
in experiment, theory, and
simulation~\cite{mueth98,ohern,coppersmith96,radjai98}.  Our data are
consistent with a small peak in $P(f)$ near the mean force or with a
plateau below the mean force.  We found that the force distributions,
$P(f)$, were well represented in all cases by the functional form
\begin{equation}
P(f)=a\,(1-b\,e^{-cf^2})\,e^{-df}
\label{eq2}
\end{equation}
which is a slight generalization of that proposed by Mueth {\em et
al}.~\cite{mueth98}.

Data was taken at both the top and bottom boundaries of the pack.  Very
little force was carried by the walls, so that the total force at the
bottom, as measured from the carbon dots, was typically within 15\% of the
force at the top.  As had been found previously \cite{mueth98}, $P(f)$ was
equivalent between the two surfaces, and the following data represent an
average of the results at the top and bottom in order to improve the
statistics.

The total forces on both boundaries were typically within 15\% of the
applied force.  This result implies that the carbon marks were dominated
by normal forces, as additional tangential forces would have increased the
total force.  The equivalence between total measured force and applied
force also suggests that there were not a significant number of duplicate
marks resulting from bead movement as these would also tend to raise the
measured total force.

\subsection{$P(f)$:  Amorphous vs.\ Crystalline}

Amorphous packs were recreated before each run, which ensured that each
run was independent of the others.  Figure~\ref{fig5}a shows $P(f)$ for
smooth amorphous packs which have been calibrated using the polynomial
method.  The data shown represent an average over 32 independent packings.
Each run gave approximately 1800 force measurements.  The distribution
agrees well with that reported by Mueth {\em et al}.~\cite{mueth98}.  The
solid line in the figure is a fit to Equation~\ref{eq2}; the fitting
parameters are shown in Table~\ref{table1}.  The error bars represent the
statistical error for multiple data sets, and the insets show
$P(f)$ for low forces calculated using both calibration interpolation
methods.  We also measured the correlations between pairs of normal
forces, and we found no significant correlations for amorphous arrangements.

$P(f)$ for multiple HCP crystals, $h=45$, of smooth beads is shown in
figure~\ref{fig5}b.  This represents an average over 26 runs, distributed
over three different HCP crystals with each run giving approximately 2200
force measurements.  These data show a similar form to that of the
amorphous packs, but there appears to be a slightly higher peak in the
smooth crystal data for forces near the average value, $f\approx1$, as
well as a slightly steeper exponential decay at large $f$.  We have seen
some evidence of aging (a gradual settling of the grains into a more
jammed state) for a single HCP crystal which had been previously subjected
to more than 40 runs.  We have excluded these data from our results.

To test the robust nature of the probability distribution, we constructed
an FCC crystal with $h=10$.  Figure~\ref{fig5}c shows $P(f)$ to be
virtually indistinguishable from that obtained for the HCP crystal.  Due
to the difficulties of constructing a crystal incommensurate with our cell,
only one FCC crystal was created.  The data represent an average of 10
runs with approximately 2000 force measurements each.

\subsection{$P(f)$:  Smooth vs.\ Rough}

\begin{figure}[p]
\centerline{\epsfxsize=8.6cm\epsfbox{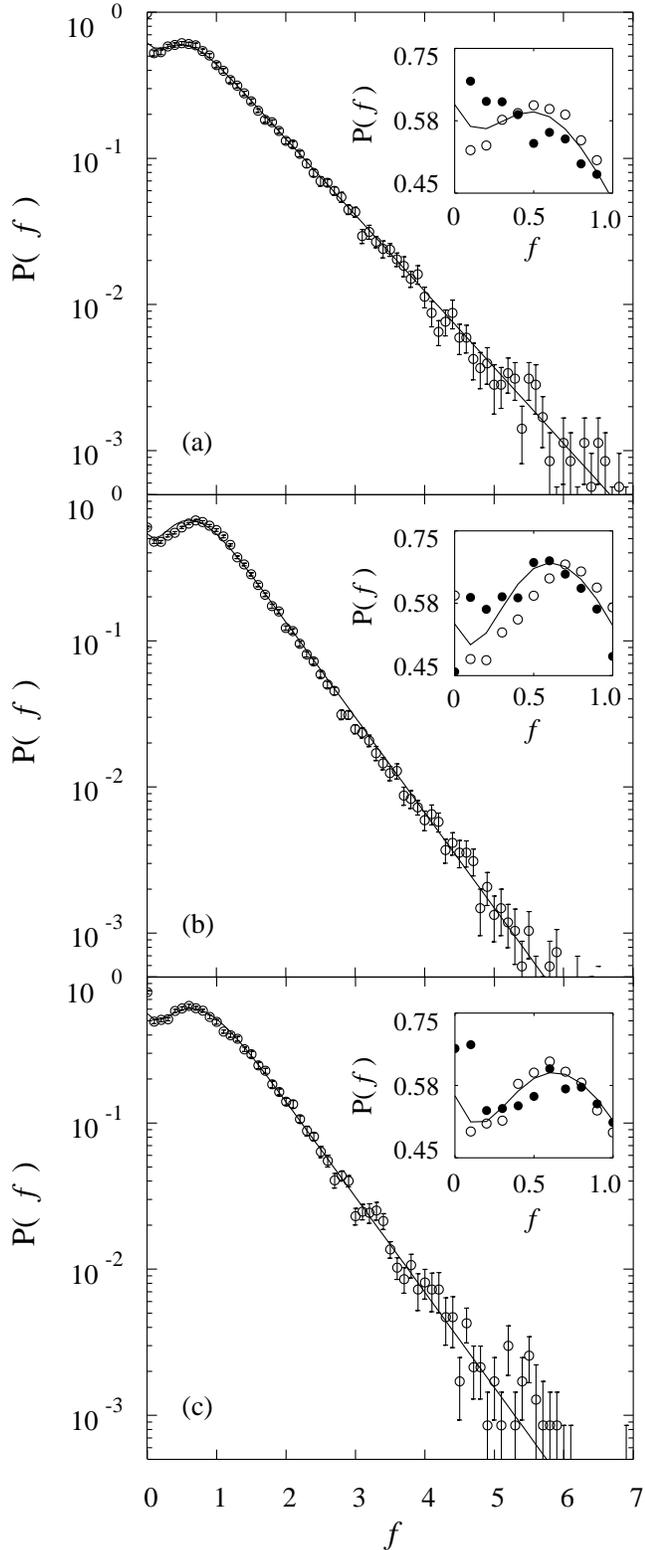}}
\vspace{.1cm}
\caption{$P(f)$ for a) amorphous packs of smooth glass spheres,  b)
HCP crystalline packs, $h=45$, of smooth glass spheres, and c) an FCC
crystalline pack, $h=10$, of smooth glass spheres.  The error bars
represent the statistical uncertainty, and the solid line represents a fit
to Eq.~\ref{eq2}.  Insets show the difference between the fourth-order
polynomial~($\circ$) and sliding parabolic~($\bullet$) interpolation
methods.}
\label{fig5}
\end{figure}

\begin{figure}[p]
\centerline{\epsfxsize=8.6cm\epsfbox{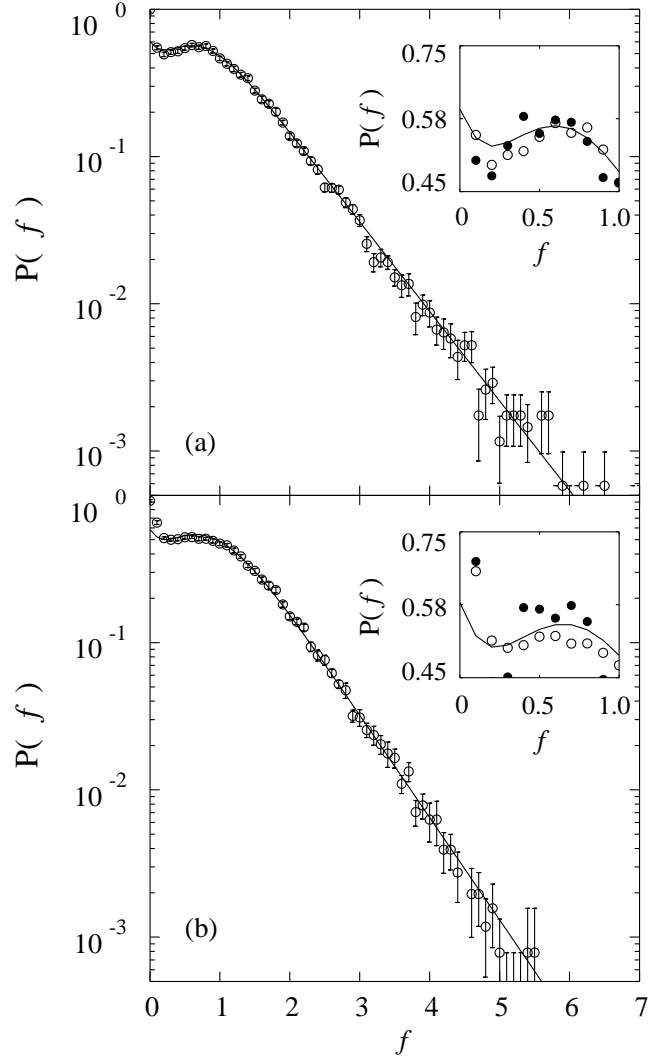}}
\vspace{.1cm}
\caption{$P(f)$ for a) amorphous packs of rough glass spheres and b) HCP
crystalline packs, $h=45$, of rough glass spheres.  The error bars
represent the statistical uncertainty, and the solid line represents a fit
to Eq.~\ref{eq2}.  Insets show the difference between the fourth-order
polynomial~($\circ$) and sliding parabolic~($\bullet$) interpolation
methods.}
\label{fig6}
\end{figure}

Amorphous and HCP crystalline packs created with the roughened beads were
also studied.  The rough amorphous data were taken from 20 independent
packings, each representing approximately 1800 force measurements.  The
rough HCP crystal data were taken from 10 data runs on one crystal with
approximately 2200 force measurements each.  The resulting force
distributions for the rough beads (Fig.~\ref{fig6}) were not 
significantly different from the distributions for the smooth beads
(Fig.~\ref{fig5}).  The fitting coefficients are given in
Table~\ref{table1}.

\begin{figure}[tb]
\centerline{\epsfxsize=8.6cm\epsfbox{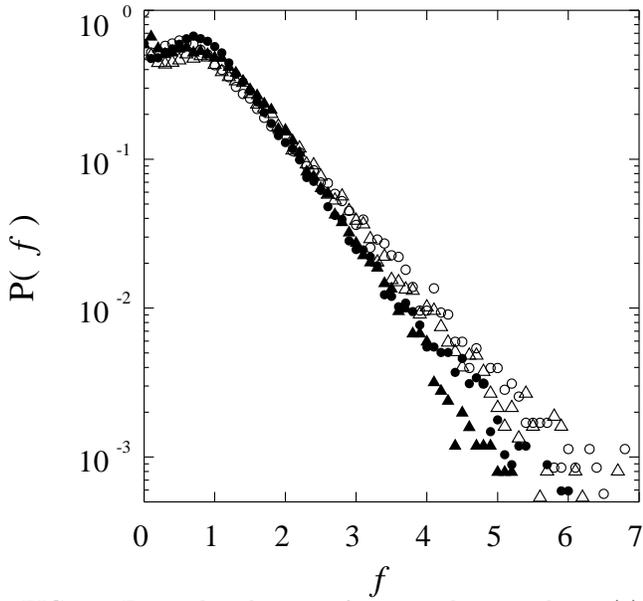}} 
\vspace{.1cm}
\caption{Force distributions for smooth amorphous~($\circ$), smooth
HCP~($\bullet$), rough amorphous~($\vartriangle$), and rough
HCP~($\blacktriangle$) granular packs.}
\label{fig7}
\end{figure}

All systems, consisting of both smooth and rough beads in both
amorphous and crystal packs,  showed similar force distributions.  The
$P(f)$ curves for four systems are shown together in Figure~\ref{fig7}.

\begin{table}[p]

\begin{tabular}{lcccc}
Fitting Parameters&  $a$&  $b$&  $c$&  $d$ \\ \hline \hline \\ 
Smooth Amorphous& 1.5& 0.59& 3.1& 1.21 \\ 
&&&&\\ 
Smooth HCP Crystal& 2.7& 0.80& 2.0& 1.50 \\ 
&&&&\\ 
Smooth FCC Crystal& 2.8& 0.80& 1.5& 1.48 \\
&&&&\\ 
Rough Amorphous& 2.4& 0.75& 1.4& 1.41 \\ 
&&&&\\ 
Rough HCP Crystal& 3.9& 0.85& 0.8& 1.60 \\
\end{tabular}
\caption{\label{table1}Values of fit parameters for the form\hfil\break
\centerline{$P(f)=a\,[1-b\,\exp(-cf^2)]\,\exp(-df)$}
}
\end{table}

\subsection{$P(f)$: Height Dependence}

\begin{figure}[tb]
\centerline{\epsfxsize=8.6cm\epsfbox{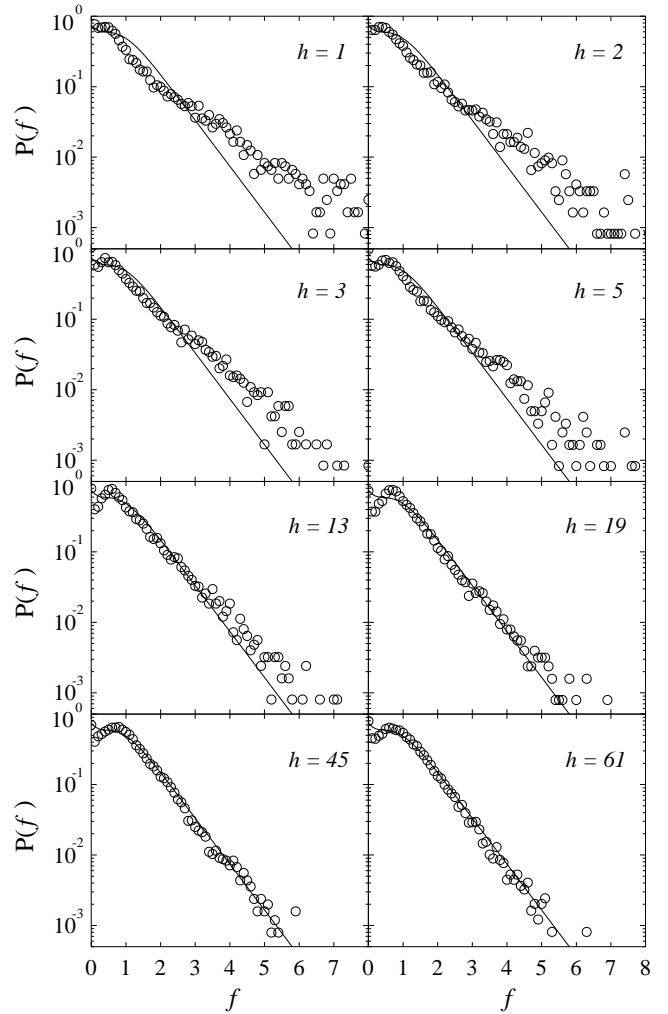}}
\vspace{.2cm}
\caption{Force distributions, $P(f)$, for smooth HCP crystals  of
various heights, $h$.  The solid line is a fit to Eq.~\ref{eq2} for the
$h=45$ smooth HCP crystal.}
\label{fig8}
\end{figure}

The aforementioned experimental results were obtained for HCP crystals
and amorphous packs which were many bead diameters in height.  We are
interested in understanding how the $P(f)$ distributions are built up as
the height is increased.  We studied this question by varying the number
of layers in HCP crystals of smooth glass beads.  Figure~\ref{fig8} shows
$P(f)$ for HCP crystals of different heights ranging from $h=1$ to $h=61$.
For comparison, the solid line in each panel represents the fit to
Equation~\ref{eq2} for the $h=45$ smooth HCP crystals.  For a single
layer, $h=1$, $P(f)$ is determined primarily by the polydispersity of the
beads themselves.  As the height is increased, $P(f)$ evolves toward a
height-independent form.  We find that $P(f)$ is robust for pack heights
greater than $h\approx15$.

\subsection{Particle Deformation}

Recent simulations on both 2-D and 3-D bead packs suggest that the
form of $P(f)$ depends on particle
deformations~\cite{makse00,thornton97,thornton98,nguyen}.  We calculate
particle deformations using a Hertzian deformation law~\cite{landau}:
\[
{\delta R\over R}=\left({2D^2\overline F^{2}\over R^{4}}\right)^{1\over3}
\]
where $\overline F$ is the mean force per bead, and $R$ is the bead radius.
$D$~is determined by material properties:
\[
D = {3(1-\sigma^2)\over2E}
\]
where $\sigma$ is Poisson's ratio and $E$ is Young's modulus.  Soda
lime glass spheres have $\sigma=0.24$ and $E=69\,\mathrm GPa$~\cite{crc}.  For
the data we have presented, $R=1.53\,\mathrm mm$, and
$\overline F\approx7.0\,\mathrm N$, resulting in particle deformations of
0.2\%.  We varied the force applied to the $h=45$ HCP crystals over a
range of $\overline F$ from 3.5\,N to 8.0\,N.  This corresponded to
deformations of 0.12\% to 0.2\%.  Over this limited range, we found that
$P(f)$ remained unchanged.

\section{Discussion and Conclusions}

Our data for $P(f)$ can be well fit by the form given in
Equation~\ref{eq2}.  For forces larger than the mean force, we found $P(f)$
to be a decaying exponential.  For forces smaller than the mean force, our
data are consistent with either a small peak or a plateau.  There was very
little change in the force distribution when the bead pack was varied from
amorphous to crystalline, although there is some evidence that the
exponential decay for large forces is slightly steeper in the crystalline
configurations.  There may also be support for an aging effect, although
this is still inconclusive.  We found $P(f)$ to be unaffected when the
inter-particle coefficient of friction was changed by a factor of three.

Packing history may also affect the results for an individual data run and
could therefore alter the form of $P(f)$.  A small number of particular
runs were found to show significantly different results, such as pure
exponential behavior.  Only after a considerable number of individual
experiments were performed and averaged could a robust probability
distribution be presented.  We find these variations in the individual runs
to be quite surprising.  They may indicate that the system is
occasionally caught in a state not representative of the average
configuration.

We do not see evidence for a change in $P(f)$ due to the deformation of
the particles over a range from 0.12\% to 0.2\%.  Simulations and
experiments performed by Makse {\em et~al}.\ suggest that $P(f)$ is a pure
exponential for small deformations, with a transition to a Gaussian
centered at $f=1$ when particle deformation is
increased~\cite{makse00}.  We estimate the deformations to vary from
0.09\% to 0.4\% in the experiments of Makse {\em et al}., while their
simulations have a much wider range.  The fact that our data
do not show a transition toward either pure exponential or Gaussian
behavior seems to be in contradiction to their experimental results.  Our
data are not necessarily inconsistent with their simulations, as our
results may be viewed as being in a transition regime between small and
large deformations.  We note, however, that simulations by
Thornton~\cite{thornton97,thornton98} as well as Nguyen and
Coppersmith~\cite{nguyen} find a slow transition toward a Gaussian profile
at high deformations (well beyond our range), but do not find a pure
exponential in any deformation regime.  At low deformations, these
simulations show a form of $P(f)$ which is qualitatively similar to our
results.  In addition, this form is also consistent with experiments by
L\o voll {\em et al}.~\cite{lovoll99} on a granular pack with no external
forcing and thus very low deformations.

We are able to make qualitative comparisons with the predictions of
some theoretical and simulation work for $P(f)$ in the $f<1$ regime.  We
find that $P(f)$ does not approach zero as $f\to0$, in contrast with the
predictions of the $q$-model~\cite{coppersmith96}.  Based on simulations,
Radjai {\em et~al}.\ report a slow power-law divergence as $f$ approaches
zero~\cite{radjai96,radjai97,radjai98}; while this form is similar to our
experimental results (since the power-law exponent is very small), it does
not account for the possibility of a peaked distribution near the mean.

Simulations performed by O'Hern {\em et~al}.~\cite{ohern} on quenched molecular
liquids in two dimensions show the most apparent similarities to our
results.  For potentials with finite repulsive terms, such as a
Lennard-Jones potential with a range cut off at its minimum value, the
$P(f)$ distribution is indistinguishable from our findings.  Also, there
may exist a correlation between granular crystals which have aged (as
mentioned above) and quenched molecular liquids at temperatures well
below the glass-transition temperature.  The
increase of the exponential slope as well as a slight peak near the mean
force in crystalline configurations are similar to those found in jammed
fluids where large forces  cannot relax when quenched below the
glass-transition temperature.

\section{Acknowledgments}
We thank Sue Coppersmith, Qiti Guo, Steve Langer, Andrea Liu, Milica
Medved, Corey O'Hern, Raghuveer Parthasarathy, Denise Sawicki, Alexei
Tkachenko, Mary Upton, and Tom Witten.  We especially thank Dan Mueth for
the background on which this work was built as well as his continued
interest, comments, and criticisms.  This work was supported by NSF under
Grant No.\ CTS-9710991 and by the MRSEC Program of the NSF under Grant
No.\ DMR-9808595.


\vspace{-0.2in} \references
\bigskip
\vspace{-0.4in}

\bibitem{jaeger96} H.~M. Jaeger, S.~R. Nagel, and R.~P. Behringer, Physics
Today {\bf 49}, 32 (1996); \rmp {\bf 68}, 1259 (1996).

\bibitem{dantu57} P.~Dantu, in {\em Proceedings of the 4th International
Conference On Soil Mechanics and Foundation Engineering} London, 1957
(Butterworths, London, 1958), Vol.\ 1, pp.\ 144-148.

\bibitem{dantu67} P.~Dantu, Ann.\ Ponts Chauss.\ {\bf IV}, 193 (1967).

\bibitem{howell97} D.~Howell, R.~P. Behringer, in {\em Powders \& Grains
97}, edited by R.~P.Behringer and J.~T. Jenkins (Balkema, Rotterdam,
1997), pp.\ 337-340.

\bibitem{ammi87} M.~Ammi, D.~Bideau, and J.~P. Troadec, J.\ Phys.\ D:
Appl.\ Phys.\ {\bf 20}, 424 (1987).

\bibitem{liu95} C.-h. Liu, S.~R. Nagel, D.~A. Schecter, S.~N. Coppersmith,
S.~Majumdar, O.~Narayan, and T.~A. Witten, Science {\bf 269}, 513 (1995).

\bibitem{mueth98} D.~M. Mueth, H.~M. Jaeger, S.~R. Nagel, \pre {\bf 57},
3164 (1998).

\bibitem{lovoll99} G.~L\o voll, K.~N. M\aa l\o y, E.~G. Flekk\o y, \pre
{\bf 60}, 5872 (1999).

\bibitem{makse00} H.~A. Makse, D.~L. Johnson, L.~M. Schwartz, \prl {\bf
84}, 4160 (2000).

\bibitem{baxter97} G.~W. Baxter, in {\em Powders \& Grains 97}, edited
by R.~P. Behringer and  J.~T. Jenkins (Balkema, Rotterdam, 1997), pp.\
345-348.

\bibitem{claudin97} P.~Claudin and J.-P. Bouchaud, \prl {\bf 78}, 231
(1997).

\bibitem{luding97} S.~Luding, \pre {\bf 55}, 4720 (1997).

\bibitem{duran98} J.~Duran, E.~Kolb, and L.~Vanel, \pre {\bf 58}, 805
(1998).

\bibitem{peraltafabi99} R.~Peralta-Fabi, C.~M\'alaga, and R.~Rechtman,
Europhys.\ Lett.\ {\bf 45}, 76 (1999).

\bibitem{ohern} C.~S. O'Hern, S.~A. Langer, A.~J. Liu, and S.~R. Nagel, in
press.  cond-mat/0005035.

\bibitem{coppersmith96} S.~N. Coppersmith, C.~Liu, S.~Majumdar,
O.~Narayan, and T.~A. Witten, \pre {\bf 53}, 4673 (1996).

\bibitem{radjai96} F.~Radjai, M.~Jean, J.~J. Moreau, and S.~Roux, \prl
{\bf 77}, 274 (1996).

\bibitem{radjai97} F.~Radjai, D.~E. Wolf, S.~Roux, M.~Jean, and J.~J.
Moreau, in {\em Powders \& Grains 97}, edited  by R.~P. Behringer and
J.~T. Jenkins (Balkema, Rotterdam, 1997), pp.\ 211-214.

\bibitem{nguyen99} M.~L. Nguyen, and S.~N. Coppersmith, \pre {\bf 59}, 5
(1999).

\bibitem{socolar98} J.~E.~S. Socolar, \pre {\bf 57}, 3204 (1998).

\bibitem{sexton99} M.~G. Sexton, J.~E.~S. Socolar, and D.~G. Schaeffer,
\pre {\bf 60}, 1999 (1999).

\bibitem{claudin98} P.~Claudin, J.-P. Bouchaud, M.~E. Cates, and J.~P.
Wittmer, \pre {\bf 57}, 4 (1998).

\bibitem{radjai98} F.~Radjai, D.~Wolf, M.~Jean, and J.~J. Moreau, \prl
{\bf 80}, 61 (1998).

\bibitem{thornton97} C. Thornton, KONA Powder and Particle {\bf 15}, 81
(1997).

\bibitem{thornton98} C. Thornton and S. J. Antony, Phil.\ Trans.\ Roy.\
Soc.\ A.\ {\bf 356}, 2763 (1998)

\bibitem{nguyen} M.~L. Nguyen, and S.~N. Coppersmith, in press.
cond-mat/0005023.

\bibitem{tkachenko99} A.~V. Tkachenko, T.~W. Witten, \pre {\bf 60}, 687
(1999).

\bibitem{eloy97} C.~Eloy and E.~Cl\'ement, J.\ Phys.\ I\ France {\bf 7},
1541 (1997).

\bibitem{thornton00} C.~Thornton, Geotechnique {\bf 50}, 43 (2000).

\bibitem{pouliquen97} O.~Pouliquen, M.~Nicolas, and P.~D. Weidman, \pre
{\bf 79}, 3640 (1997).

\bibitem{materials} We used Super Nu-Kote SNK-11 1/2 carbon paper and
Hammermill Laser Print Long Grain Radiant White paper.

\bibitem{delyon90} F.~Delyon, D.~Dufresne, and Y.-E. L\'evy, Ann.\ Ponts
Chauss.\ 22 (1990).

\bibitem{landau} L.~Landau and E.~Lifshitz, {\em Theory of Elasticity}
(Pergamon, Oxford, 1959), pp.\ 34-35.

\bibitem{crc} {\em CRC Handbook of Tables for Applied Engineering
Science}, edited by R. Bolz and G. Tuve, (CRC Press, Boca Raton, 1970),
pp.\ 160-163.
\end{document}